\documentclass[aip,jcp,twocolumn,superscriptaddress,numerical,reprint]{revtex4}
\usepackage{graphicx}
\usepackage{amssymb}
\usepackage{citesort}

\begin{document}
\title{
\bf Electron Correlations and Two-Photon States in Polycyclic Aromatic Hydrocarbon Molecules: A Peculiar Role of Geometry}
\author{Karan Aryanpour}
\affiliation{Department of Physics, University of Arizona, Tucson, AZ 85721}
\author{Alok Shukla}
\affiliation{Department of Physics, Indian Institute of Technology, Powai, Mumbai - 400076, India.}
\author{Sumit Mazumdar}
\affiliation{Department of Physics, University of Arizona, Tucson, AZ 85721}
\affiliation{College of Optical Sciences, University of Arizona, Tucson, AZ 85721}
\date{\today}
\begin{abstract}
{We present numerical studies of one- and two-photon excited states ordering in a number of polycyclic aromatic hydrocarbon molecules: coronene, hexa-peri-hexabenzocoronene and circumcoronene, all possessing $D_{6h}$ point group symmetry versus ovalene with $D_{2h}$ symmetry, within the Pariser-Parr-Pople model of interacting $\pi$-electrons.
The calculated energies of the two-photon states as well as their
relative two-photon absorption cross-sections within the interacting model are qualitatively different from single-particle descriptions.
More remarkably, a peculiar role of molecular geometry is found. The consequence of electron correlations is far stronger for ovalene, 
where the lowest spin-singlet two-photon state is a quantum superposition of pairs of lowest spin triplet states, 
as in the linear polyenes. The same is not true for $D_{6h}$ group hydrocarbons. Our work indicates significant covalent character, in valence bond
language, of the ground state, the lowest spin triplet state and a few of the lowest two-photon states in $D_{2h}$ ovalene but not in 
those with $D_{6h}$ symmetry.}
\end{abstract}
\maketitle
\section{Introduction and Background}
\label{sec:intro}
\par The role of electron correlations on the photophysics of $\pi$-conjugated systems has been of continuing interest. 
Existing research has largely focused on the effect of electron correlations on one-photon allowed optical states.
These states can be described nearly quantitatively upon including the configuration interaction (CI) between 
many-electron configurations that are singly excited from the mean-field ground state. We will refer to these configurations as
one electron-one hole (1e-1h) excitations hereafter. Such single CI (hereafter SCI) studies \cite{Salem66a}
and equivalent approaches (time-dependent density functional theory, Bethe-Salpeter equation)
have been widely employed to understand one-photon excitations.
The dominant effect of the SCI in finite molecules is to
remove the degeneracies between 1e-1h excitations that are obtained within the one-electron or mean-field theories \cite{Salem66a}.
The effect on extended systems is similar, with electron-hole interactions
leading to exciton states separated from a continuum band 
\cite{Abe92a,Chandross94a,Rohlfing99a,Horst99a,Puschnig02a,Ruini02a,Igumenshchev07a,Ando97a,Spataru04a,Chang04a,Perebeinos04a,Zhao04a}. 
There exists a wide body of experimental literature that supports these theoretical concepts in
$\pi$-conjugated systems, including in extended carbon nanostructures \cite{Sariciftci97a,Vardeny09a,Jorio07a}.
\par In systems with inversion symmetry spin-singlet excited states accessible by one- and two-photon excitations 
from the singlet ground state are distinct. The consequences of electron correlations on the one-photon forbidden but
two-photon allowed states in correlated-electron systems are often more intricate. Two-photon states, even at the lowest energies,
can have significant contributions from two electron-two hole (2e-2h) excitations, which have
CI with both the correlated ground state and up to four electron-four hole (4e-4h) excitations, as was shown \cite{Tavan87a} within the Pariser-Parr-Pople
(PPP) $\pi$-electron Hamiltonian \cite{Pariser53a,Pople53a}. The higher order CI can lower
the energy of the two-photon states significantly relative to the one-photon states. 
Thus at the mean-field or SCI levels of approximation, the lowest excited state in linear
polyenes is the odd-parity optical 1$^1$B$_u^+$ state, with the lowest even parity 2$^1$A$_{g}^-$ state occurring above it 
(the superscript $1$ indicates singlet spin state and the ``plus'' and ``minus'' superscripts refer to charge-conjugation symmetry, hereafter CCS). This
is the ``normal'' excited state ordering. Because of the higher order CI, however, in long polyenes \cite{Hudson82a,Kohler88a} as well as in the 
polydiacetylenes \cite{Lawrence94a}, the two-photon 2$^1$A$_g^-$ occurs {\it below} the 1$^1$B$_u^+$. We will refer to this 
latter ordering as ``reversed'' excited state ordering. The energy of the lowest two-photon states, relative to that of the lowest one-photon optical
state, as well as the extent of 2e-2h character of the lowest two-photon states, are then semiquantitative {\it ``measures''} of electron correlation effects.
\par Reversed excited state ordering is best understood within valence bond (VB) concepts, within which the ground state as well as the 
lowest spin-singlet even parity two-photon states
are predominantly covalent \cite{Hudson82a,Soos84a,Ramasesha84a,McWilliams91a,Baeriswyl92a}. The dipole selection rule limits optical excitations from
the even parity covalent ground state to odd parity ionic states, which are necessarily higher in energy than covalent
states. Within VB theory it is also possible to explain naturally the quantum-entangled two-triplet character of the
2$^1$A$_g^-$ in polyenes, whose energy is almost exactly twice the energy of the lowest spin triplet 1$^3$B$_u$ (we suppress the charge-conjugation symmetries for triplets as we do not calculate triplet absorption spectra in this work). 
Actually, in long polyenes, several of the low-lying $^1$A$_g^-$ states are quantum superpositions of low-lying triplets, 
which are themselves also covalent within the VB language \cite{Tavan87a}. Indeed, within the Hubbard model description of
linear chains with very large onsite Coulomb interactions, the lowest spin-triplet and singlets are both spin excitations, while 
the optical state is necessarily a charge-excitation \cite{Ovchinnikov70a}. The extent to which higher energy
two-photon states are of ``two-triplet'' character, and the separation between them in energy space, are therefore additional measures of electron 
correlation effects: the smaller the separation between spin singlet states with two-triplet character, the stronger is the electron correlation effect.
\par Although the original understanding of reversed excited state ordering came from PPP-model calculations, over the years there have been
steady efforts to develop more sophisticated correlated {\it ab initio} approaches to reversed excited state ordering in polyenes. Quantum chemical
approaches that reproduce the PPP results for polyenes include the complete active space second order perturbation theory (CASPT2) and
third-order coupled cluster theory \cite{Schreiber08a,Silva10a}, and the extended algebraic
diagrammatic construction (ADC(2)-x) method \cite{Starcke06a,Knippenberg12a,Krauter13a}. Other methods that have been applied are density 
functional theory-based multiple reference configuration interaction (DFT-MRCI) \cite{Silva08b}, and the OM2-MRCI \cite{Schmidt12a}. 
All of these approaches have confirmed the 2e-2h character of the lowest two-photon states in the polyenes, and except for the absence of CCS 
symmetry within the {\it ab initio} theories, no significant difference has been found between the results obtained by methods which incorporate 
a high degree of correlations \cite{Schmidt12a}. On the other hand, because of the dominant double-excitation character
of the two-photon states, the number of many-electron configurations that need to be retained in the calculations increases rapidly with system
size, and calculations of polyenes within the {\it ab initio} approaches are limited to relatively small system sizes (between 8 to 12 carbon atoms
usually). The rapid increase in the number of multiply excited configurations with system size has precluded high order CI studies for large
molecules starting from {\it ab initio} approaches. Accurate studies of two-photon states and excited state ordering in large
$\pi$-conjugated systems have therefore been done mostly within the semiempirical PPP $\pi$-electron-only model
Hamiltonian. Density matrix renormalization group (DMRG) calculations have been performed
for long chain systems \cite{Ramasesha00a,Race03a,Barcza13a} within the PPP Hamiltonian. 
Ref.~\onlinecite{Barcza13a} in particular was able to obtain several close-lying $^1$A$_g^-$ states below the 
1$^1$B$_u^+$ in the long chain limit of polydiacetylenes, that were at nearly twice the energies of 
the lowest triplets. The results for the two-photon states in the long chain limit here are in excellent agreement with experiments \cite{Lawrence94a}.
Normal, as opposed to reversed excited state ordering occurs in $\pi$-conjugated polymers 
with large unit cells containing phenyl groups is found within PPP calculations, and the mechanism of this is also well understood \cite{Soos93a,Chandross99a}. 
The optical 1$^1$B$_u^+$
state has contributions from Frenkel intraunit 1e-1h excitations as well as interunit charge-transfer excitations, with the charge-transfer
contribution dominating in linear polyenes \cite{Chandross99a}. The 2$^1$A$_g^-$ is reached from the 
1$^1$B$_u^+$, by a second interunit charge-transfer
in the reverse direction \cite{Chandross99a}. As the charge-transfer contribution to the 
1$^1$B$_u^+$ decreases and the Frenkel contribution
increases, the relative energy of the 2$^1$A$_g^-$ is enhanced
and normal excited state ordering is recovered \cite{Soos93a,Chandross99a}. 
The study of excited state ordering in the low energy region of 1D carbon-based systems can be taken to be complete.
\par In contrast to the 1D systems, theoretical and experimental studies of two-photon states and excited state ordering in 
two-dimensional (2D) polycyclic aromatic hydrocarbons (PAHs) are in their early stages. 
We have very recently presented joint theory-experiment results of investigations for  
a few select PAH molecules \cite{Aryanpour14a}. The molecules investigated theoretically in all cases had 
$D_{6h}$ point group symmetry. The reason for choosing this particular set was because the two molecules investigated
experimentally, coronene (C$_{24}$H$_{12}$), which is commercially available, and hexa-peri-hexabenzocoronene (C$_{42}$H$_{18}$, hereafter HBC), which was synthesized by our experimental colleagues, both had $D_{6h}$ symmetry.
Correlation effects in these $D_{6h}$ PAH molecules were shown to be significant, and
reversed excited state ordering was demonstrated experimentally and theoretically 
in both coronene and HBC. As per our measures discussed above, correlation effects in PAHs are stronger than in the $\pi$-conjugated
polymers with phenyl groups, in spite of the larger one-electron bandwidths of the former. At the same time, 
they are weaker in PAHs than in linear polyenes, since in PAHs
the lowest two-photon states were found from our calculations to be predominantly (though not
entirely) 1e-1h in character, in contrast to the 2$^1$A$_g^-$ in the polyenes, in which the 2e-2h contributions
dominate over the 1e-1h contributions \cite{Aryanpour14a}.
 The weaker correlation effect in PAHs, relative to polyenes, is expected because of the larger
one-electron bandwidth of PAHs in which each carbon (C) atom has three as opposed to two neighboring C-atoms.

\par
\begin{figure*}
\includegraphics[width=7.0in]{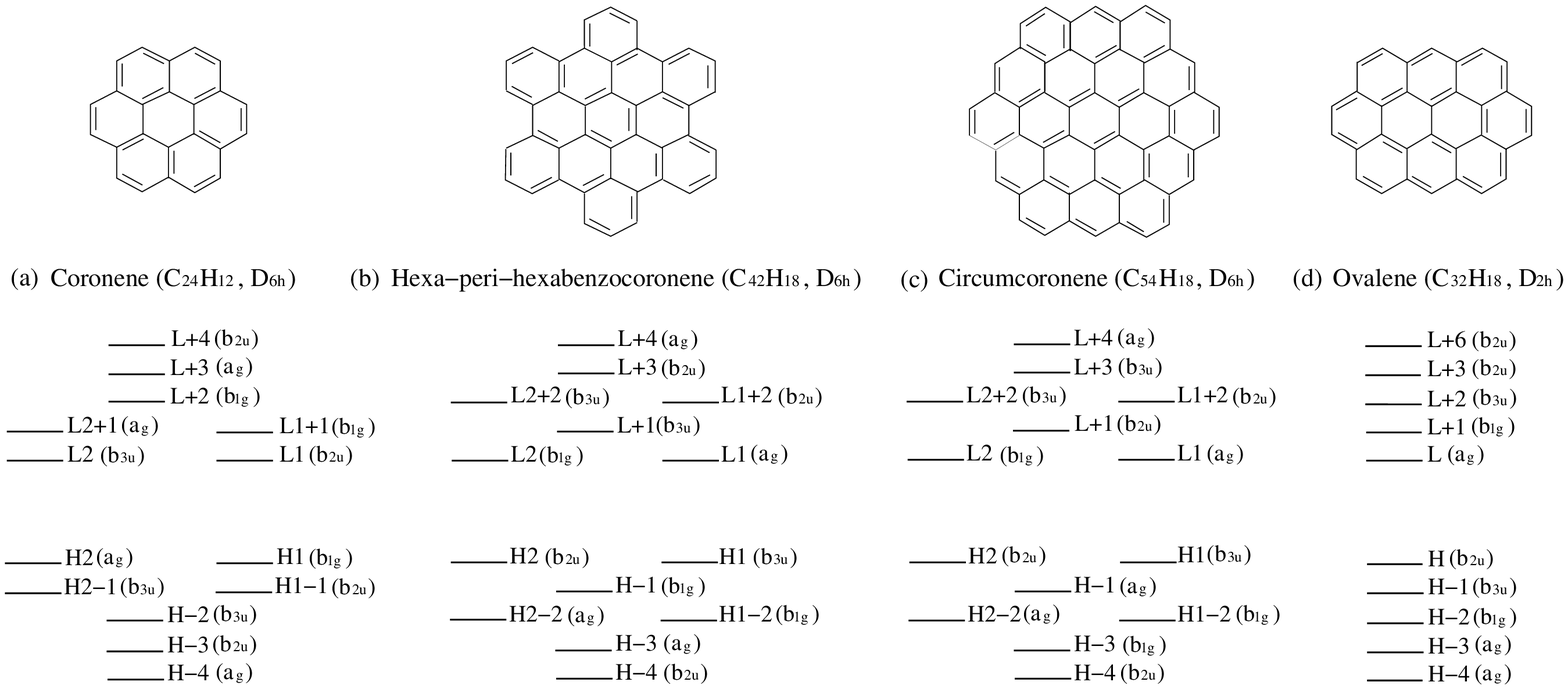}
\caption{(a) Coronene (C$_{24}$H$_{12}$), (b) HBC (C$_{42}$H$_{18}$), (c) circumcoronene (C$_{54}$H$_{18}$) and (d) ovalene. Schematics of the
frontier HF MOs 
that occur within the energy space between and including HOMO$-$4 (H$-$4) to LUMO$+$4 (L$+$4) are given. The $D_{2h}$ orbital symmetry is 
specified next to each MO. The PPP-MRSDCI calculations were done for the full MO-space of coronene, and ovalene; for HBC and circumcoronene 
the outermost 5 and 8 pairs of MOs, $\pm$ 4.62 and $\pm$ 4.54 eV away from the chemical potential, respectively, were ignored in our calculation.} 
\label{molecules}
\end{figure*}

 We have subsequently performed high order CI calculations of higher energy one- versus two-photon singlet excited states, 
and spin triplet states of these $D_{6h}$ PAH molecules to get a more complete picture. We have also implemented the same calculations
for a different PAH molecule possessing $D_{2h}$ point group symmetry,  
ovalene (C$_{32}$H$_{18}$), for comparison (see Fig.~\ref{molecules}). In addition to calculating energies we have also performed
detailed wavefunction analyses.
Our computational results lead to a surprising conclusion, viz., as measured against our criteria above, 
{\it the extent of electron correlation effects in PAHs is geometry-dependent!} 
Correlation effects are considerably stronger in $D_{2h}$ ovalene than in the $D_{6h}$ PAH molecules with comparable or even 
slightly larger sizes. A ``hierarchy'' of correlation effects (polyenes $>$ $D_{2h}$ PAHs $>$ $D_{6h}$ PAHs) thus emerges, which
has no obvious explanation. We emphasize that the effect we demonstrate in this work is separate from that of the edges which drives the 
difference between graphene nanoribbons with zigzag
versus armchair edges \cite{Guclu10a}, as two of the three $D_{6h}$ PAH molecules we study, as well as ovalene with $D_{2h}$ symmetry,
have zigzag edges. Difference is also known to exist between finite triangular versus hexagonal graphene fragments with zigzag edges 
\cite{Heiskanen08a,Zarenia11a}. In the former
there always occur degenerate one-electron levels at the chemical potential, as a consequence of which the ground state wavefunction
of the neutral molecule is
a superposition of degenerate configurations. In contrast, there occurs a gap in the one-electron levels of small hexagonal fragments, and the closed shell
ground state wavefunction is nondegenerate. In the present case, the overall ground state wavefunction is nondegenerate in all cases, and the
difference we find cannot be understood within one-electron theory.
\section{Theoretical Model and Computational Method}
\label{sec:model}
\par The PPP Hamiltonian \cite{Pariser53a,Pople53a} is written as,
\begin{eqnarray}
\label{PPP_Ham}
 H_{PPP}=-t\sum_{\langle ij \rangle\sigma}
(\hat{c}_{i\sigma}^\dagger \hat{c}_{j\sigma}+\hat{c}_{j\sigma}^\dagger \hat{c}_{i\sigma}) + U\sum_{i} \hat{n}_{i\uparrow} \hat{n}_{i\downarrow} + \nonumber \\
 \sum_{i<j} V_{ij} (\hat{n}_{i}-1)(\hat{n}_{j}-1)\,,\hspace{0.95in}
\end{eqnarray}
where $\hat{c}^{\dagger}_{i\sigma}$ creates a $\pi$-electron of spin $\sigma$ on carbon atom $i$, 
$\hat{n}_{i\sigma} = \hat{c}^{\dagger}_{i\sigma}\hat{c}_{i\sigma}$ is the number of electrons of spin $\sigma$ on atom $i$, 
and $\hat{n}_i=\sum_{\sigma} \hat{n}_{i\sigma}$
The one-electron hopping integral $t$ is between nearest neighbor carbon atoms $i$ and $j$, 
$U$ is the Hubbard repulsion between two electrons occupying the same atomic $p_z$ orbital,
and $V_{ij}$ is the long range intersite Coulomb interaction. Given the semiempirical nature of the PPP Hamiltonian, in the context of the present work,
electron correlations effects are the ones which are not captured in the
tight-binding theory and are included by performing configuration interaction calculations. For the correlated-electron calculations reported below we choose standard $t=2.4$ eV \cite{Ramasesha84a,Tavan87a,Baeriswyl92a}, and 
obtain $V_{ij}$ from 
the parameterization \cite{Chandross97a}, 
$V_{ij}=U/\kappa\sqrt{1+0.6117 R_{ij}^2}$, where $R_{ij}$ is the distance in $\mathring{\textrm{A}}$
between carbon atoms $i$ and $j$ and $\kappa$ is an effective dielectric constant. The restriction of electron
hopping to nearest neighbors preserves alternancy symmetry.
We further choose $U=8$ eV and $\kappa=2$. These parameters have given excellent fits previously to the 
singlet excited states and spectra of poly-paraphenylenevinylene \cite{Chandross97a}, polyacenes \cite{Sony07a},
single-walled carbon nanotubes \cite{Wang06a}, coronene and HBC (in the case of coronene, quantitative
agreement was obtained also between the calculated and experimental energies of the lowest triplet \cite{Aryanpour14a}).
\par The top panel of Fig.~\ref{molecules} presents the PAH molecules studied in this work and the point group symmetry they possess. 
 Since the calculations of the lowest energy states have already been reported for these $D_{6h}$ PAH molecules in Ref.~\onlinecite{Aryanpour14a}, explicit comparisons 
of theoretical one- and two-photon absorption spectra for ovalene are made only against coronene as the 
representative of PAH molecules with $D_{6h}$ symmetry. However, more detailed comparisons of spin-triplet and higher energy two-photon states,
and in particular, the wavefunctions of the latter, are
made against all the $D_{6h}$ PAH
molecules of Fig.~\ref{molecules}. Our calculations were done with fixed C-C bond lengths of
$1.4$ $\mathring{\textrm{A}}$ and all bond angles of 120$^{\circ}$. 
The actual bond lengths and angles may deviate from these mean values, but the deviations are small because of the
aromatic nature of the molecules. In the case of coronene we have previously shown from explicit calculations that 
the consequences of the deviations from idealized geometry on the energies of the one- and two-photon states, and of the lowest
triplet, are negligible \cite{Aryanpour14a}.
\par One-photon states for molecules with $D_{6h}$ symmetry belong to the doubly degenerate $^1$E$_{1u}^+$ symmetry subspace, while two-photon
states belong to either $^1$A$_{1g}^-$ or $^1$E$_{2g}^-$ symmetry subspaces. 
As $D_{6h}$ is not an Abelian group, we have adopted the $D_{2h}$ nomenclature also for
discussions of the excited states in the $D_{6h}$ PAHs. Consistent adoption of the $D_{2h}$ point group symmetry for 
all cases of interest allows one-to-one transparent comparisons between the $D_{6h}$ PAH molecules of Figs.~\ref{molecules}(a)-(c) and ovalene, 
which is of $D_{2h}$ symmetry.
The ground state is then 1$^1$A$_g^-$ in all cases.
The dipole selection rule dictates that one-photon (two-photon) absorption occur to states with
inversion and charge-conjugation symmetries opposite to (same as) that of the ground state. Hence the allowed
one-photon (two-photon) states belong to the $^1$B$_{2u}^+$ and $^1$B$_{3u}^+$ ($^1$A$_{g}^-$ and $^1$B$_{1g}^-$)
symmetry subspaces. In addition,
PAHs can also have $^1$B$_{2u}^-$ and  $^1$B$_{3u}^-$ states whose spatial and charge-conjugation symmetries preclude
both one- and two-photon absorptions.
The lowest two triplet states for PAH molecules in Figs.~\ref{molecules}(a)-(d) are 1$^3$B$_{2u}$ and 1$^3$B$_{3u}$ states.
\par Our calculations were done using the multiple reference singles and doubles configuration interaction (MRSDCI) 
which is a powerful many-body approach for incorporating CI between the dominant excitations up to quadruples. 
CI calculations are done in the basis of excitations from the ground state, in which electrons fill up the lowest energy molecular orbitals (MOs), 
obtained as self-consistent restricted Hartree-Fock (HF) solutions
of the PPP Hamiltonian in Eq.~\ref{PPP_Ham}. 
The lower panel of Fig.~\ref{molecules} depicts a number of frontier MOs around the HF gap in  
coronene, HBC, circumcoronene and ovalene with their corresponding $D_{2h}$ orbital symmetries. HOMO and LUMO stand for 
the highest occupied and lowest unoccupied MOs, respectively in the HF ground state. 
In the first three PAH molecules (Figs.1(a)-(c)), $D_{6h}$ symmetry results in two-fold degeneracy of the HOMO (LUMO) depicted as H1 and H2 (L1 and L2) and 
also some other lower (higher) frontier MOs while true $D_{2h}$ symmetry gives non-degenerate frontier MOs as is seen in Fig.~\ref{molecules}(d) for ovalene. 
The MRSDCI procedure is well known
and has been described in detail in our earlier work \cite{Aryanpour14a}. 
Suffice it is to say that we obtain quantitative fits to the energies of the
optical singlet state and the lowest spin-triplet state for coronene within the PPP-MRSDCI calculations. In Table~\ref{t1} we have 
given the dimensions
of the final Hamiltonian matrices that were used to calculate energies and wavefunctions iteratively for the excited states
of the PAH molecules within different symmetry subspaces. 
\par Optical transitions between the ground and excited states are obtained from the matrix elements 
between these eigenstates of the transition dipole operator 
${\bf{\hat\mu}}=\sum_{i}{\bf r_i}{\hat n}_i$
with ${\bf r_i}$ the position vector associated with C-atom $i$ (we work in reduced units with the electronic charge $e=1$). 
The frequency-dependent one-photon absorption coefficient $\alpha(\omega)$ is given by,        
\begin{eqnarray}
\label{1ph-absorp}
 \alpha(\omega)=\omega~{\textrm{Im}}\Big[\sum_{n}\big(
 \frac{\mu_{x,gn}^2}{E_{n}-\hbar \omega-i\delta}+\frac{\mu_{y,gn}^2}{E_{n}-\hbar \omega-i\delta}\big)\Big]\,,
\end{eqnarray}
where $\mu_{x,gn}=\langle G|\hat{\mu}_{x}|{\textrm n}^1{\textrm B}_{3u}^+\rangle$ and $\mu_{y,gn}=\langle G|\hat{\mu}_{y}|{\textrm n}^1{\textrm B}_{2u}^+\rangle$
with $|G\rangle$ the ground state, $E_{n}$ the energies of $|{\textrm n}^1{\textrm B}_{2u}^+\rangle$  and $|{\textrm n}^1{\textrm B}_{3u}^+\rangle$ virtual one-photon states and $\delta$ the energy 
linewidth (set to 0.03 eV in our calculations).  
We also evaluate the two-photon transition moments \cite{McWilliams91a}
\begin{equation}
\label{2ph-absorp}
M_{\lambda\lambda^{\prime}}(f)=\sum_{n}\frac{\mu_{\lambda,gn}~\mu_{\lambda^{\prime},nf}}{E_{n}-(\frac{E_{f}}{2})}\,, 
\end{equation}
where $\lambda$, $\lambda^{\prime}=x,y$, and
\begin{eqnarray}
\mu_{x,nf}=\langle {\textrm n}^1{\textrm B}_{3u}^+|\hat{\mu}_{x}|{\textrm f}^1{\textrm A}_{g}^-\rangle~~~{\textrm {or}}~~~\langle {\textrm n}^1{\textrm B}_{2u}^+|\hat{\mu}_{x}|{\textrm f}^1{\textrm B}_{1g}^-\rangle \nonumber \\
\mu_{y,nf}=\langle {\textrm n}^1{\textrm B}_{2u}^+|\hat{\mu}_{y}|{\textrm f}^1{\textrm A}_{g}^-\rangle~~~{\textrm {or}}~~~\langle {\textrm n}^1{\textrm B}_{3u}^+|\hat{\mu}_{y}|{\textrm f}^1{\textrm B}_{1g}^-\rangle\,.
\end{eqnarray}
In the above $|{\textrm f}^1{\textrm A}_{g}^-\rangle$ and $|{\textrm f}^1{\textrm B}_{1g}^-\rangle$ 
are the final two-photon states.
The two-photon absorption cross-section, hereafter TPA, has multiple components, whose strengths are given by \cite{Boyd92a},
\begin{eqnarray}
\label{2ph-intense}
\text{TPA}_{1111}(f) = |M_{xx}(f)|^2,~~~\text{TPA}_{2222}(f) = |M_{yy}(f)|^2,\nonumber \\ 
\text{TPA}_{1212}(f) = |\frac{1}{2}(M_{xy}(f)+M_{yx}(f))|^2\,.\hspace{1.2cm}
\end{eqnarray}
The final states in the $1111$ and $2222$ components of two-photon absorption are necessarily 
of $^1$A$_{g}^-$ symmetry, while for the $1212$ component the final state symmetry is $^1$B$_{1g}^-$. 
\begin{table}
{\footnotesize
\caption{Dimensions of the final PPP-MRSDCI matrices for the different $D_{2h}$ symmetry subspaces of the PAH molecules in 
Figs.~\ref{molecules}(a)-(d). The final MRSDCI wavefunctions contained basis functions with coefficients as 
small as 0.06 for circumcoronene and 0.04 for the other molecules.}
\label{t1}
\begin{tabular}{l}
\hline \hline 
\\ 
\hspace{1.1cm}Coronene\hspace{0.6cm}Ovalene\hspace{1.0cm}HBC\hspace{0.6cm}Circumcoronene  \\ \\
$^1$A$_{g}$:\hspace{0.55cm}$2045687$\hspace{0.8cm}$3231505$\hspace{0.8cm}$3371103$\hspace{0.8cm}$3864837$  \\ \\
$^1$B$_{3u}$:\hspace{0.45cm}$972754$\hspace{0.9cm}$2744346$\hspace{0.8cm}$1719854$\hspace{0.8cm}$3133234$  \\ \\
$^1$B$_{2u}$:\hspace{0.4cm}$1082466$\hspace{0.8cm}$3385422$\hspace{0.8cm}$2227463$\hspace{0.8cm}$3645309$  \\ \\
$^1$B$_{1g}$:\hspace{0.4cm}$1162244$\hspace{0.8cm}$3348726$\hspace{0.8cm}$3167504$\hspace{0.8cm}$3745386$ \\ \\
$^3$B$_{3u}$:\hspace{0.4cm}$1495328$\hspace{0.8cm}$3836862$\hspace{0.8cm}$1709000$\hspace{0.8cm}$3027448$  \\ \\
$^3$B$_{2u}$:\hspace{0.4cm}$2902834$\hspace{0.8cm}$4073605$\hspace{0.8cm}$1885670$\hspace{0.8cm}$3904921$  \\ \\
\hline \hline
\end{tabular}
}
\end{table}
\section{Results and Analyses}
\label{sec:results}
\par\noindent{\it Two-photon absorptions: H\"uckel model}. In order to have a clear understanding of electron correlation effects on the photophysics
of PAHs, it is useful to first determine the one-photon and two-photon absorptions within the noninteracting
H\"uckel model ($U=V_{ij}=0$ limit of the PPP model in Eq.~\ref{PPP_Ham}). We have calculated the one-photon absorption $\alpha(\omega)$ and the
two-photon absorptions TPA$_{1111}$, TPA$_{2222}$ and TPA$_{1212}$ in
both coronene and ovalene. 
Coronene has been studied previously in our earlier work \cite{Aryanpour14a} where its one- and two-photon excited states were both experimentally and 
theoretically determined and discussed in detail. Here we repeat the calculations and present the results for coronene as a representative of the
$D_{6h}$ group PAHs solely for the sake of comparison with ovalene possessing $D_{2h}$ symmetry.

The hopping integral $t$ sets all energy scales within the H\"uckel model. One then has a choice of using realistic $t=2.4$ eV,
or a value that will fit the experimental ground state absorption energy (note that the ratio E(2$^1$A$_g^-$)/E(1$^1$B$_{2u}^+$), where E($\cdots)$ is the energy
of the state $\cdots$, all other energy ratios, and the relative intensities of one and two-photon absorptions are all independent of $t$). Our calculations are for $t=3.8$ eV, as this value of
$t$ reproduces the experimental energy of the allowed optical absorption in coronene (4.1 eV)
\cite{Aryanpour14a}. 
We have used the same $t$ also for ovalene,
as the energy of the dominant one-photon state within the H\"uckel model in this case is the same as that obtained within the PPP model
with realistic $t=2.4$ eV (see below).
\par In Fig~\ref{Huckel}(a) we show the H\"uckel ground state absorption and the transition moments corresponding to TPA$_{1111}$ and TPA$_{1212}$
for coronene. The one-photon absorption is due to the four-fold degenerate transitions H1$\to$ L1,~L2 and H2$\to$L1,~L2 (see Fig.~\ref{molecules}(a)). 
The two-photon $^1$A$_g^-$ and $^1$B$_{1g}^-$ states are also degenerate within the H\"uckel limit. We have labeled these degenerate absorptions as simply
1-ph and 2-ph in the Figure, rather than giving the symmetry classifications of the final states. The two-photon states correspond to 1e-1h excitations 
whose energies are higher than the HOMO $\to$ LUMO excitations. 
The TPA$_{2222}$ is not shown separately, as the strength of this
component is nearly the same as for TPA$_{1111}$ 
within the $D_{6h}$ symmetry.
There occur only two two-photon states below 7 eV, 
the upper energy limit for our absorption spectrum analysis. Importantly, the lowest two-photon state lies 1.75 eV higher in 
energy than the optical one-photon state. 
\par Due to the lower $D_{2h}$ symmetry in ovalene, lowest dipole-allowed one-photon excitations are nondegenerate and correspond to the transitions 
(i) H$\rightarrow$L, (ii) H$-$1$\rightarrow$L$+$1, and (iii) the degenerate pair H$-$1$\rightarrow$L and H$\rightarrow$L$+$1 (see Fig.~\ref{molecules}(d)). These transitions, along with the two-photon transition moments are shown in Fig.~\ref{Huckel}(b). 
The $1111$ and $1212$ components are no longer degenerate and the $2222$ component, while still degenerate with the $1111$ component, 
has much higher intensity now. The lowest two-photon state in Fig.~\ref{Huckel}(b) again lies significantly above the lowest one-photon state, by 
1.65 eV. It is also higher in energy than the dominant one-photon state at 3.56 eV by 0.62 eV.
\begin{figure*}
\includegraphics[width=7.0in]{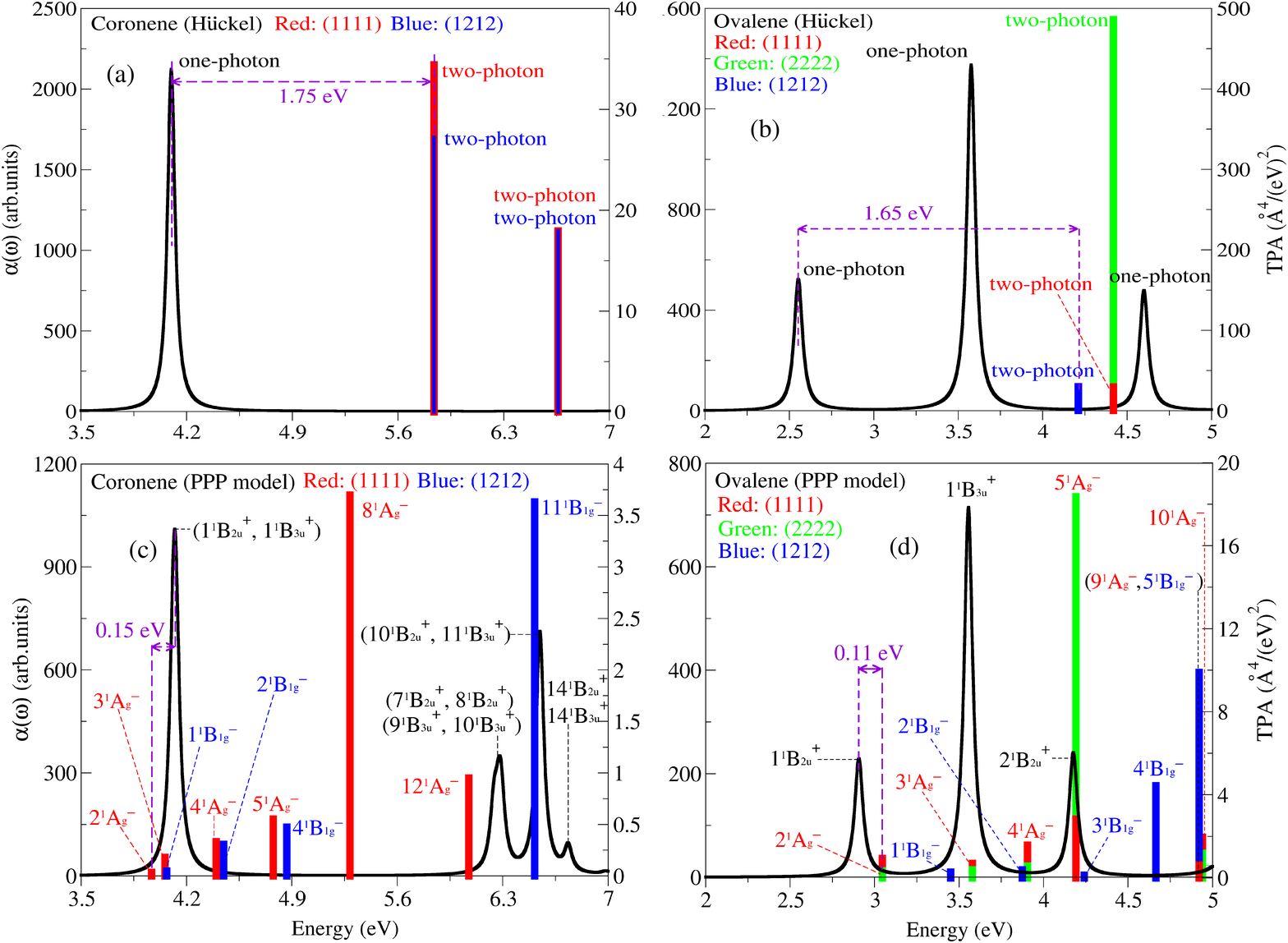}
\caption{ (a) Calculated one- and two-photon absorptions in coronene within the H\"uckel model. Red and blue colors correspond to $1111$ and $1212$ 
of the TPA components, respectively (b) The same as (a) for ovalene. The $2222$ TPA component is now included and is shown in green. 
(c) Calculated  one- and two-photon absorptions for coronene. (d) The same as (c) for ovalene. One- and two-photon excited states in (c) and (d) are 
labeled using the $D_{2h}$ point group symmetry, along with their positive (negative) charge-conjugation symmetries (TPA results assume reduced units with the electronic charge $e=1$).}
\label{Huckel}
\end{figure*}
\par\noindent{\it Two-photon absorptions: PPP-MRSDCI results.} Fig.~\ref{Huckel}(c) shows the one- and two-photon absorptions for coronene, calculated within the PPP-MRSDCI approach, with realistic $t=2.4$
eV. Electron correlations remove the four-fold degeneracies of the one-photon excitations, and the correlated 
eigenstates are superpositions
of the nointeracting excitations. The two superpositions with relative minus signs between equivalent excitations
constitute the $1^1$B$_{2u}^-$ and $1^1$B$_{3u}^-$, and are optically dark. The remaining two superpositions, with relative
plus signs between the equivalent excitations are the optically allowed degenerate $1^1$B$_{2u}^+$ and $1^1$B$_{3u}^+$, within the
$D_{2h}$ classification scheme. As seen in Fig.~\ref{Huckel}(c), the two-photon $^1$A$_{g}^-$ and $^1$B$_{1g}^-$ are nondegenerate now.
More importantly, the lowest two-photon states, the $2^1$A$_{g}^-$, $3^1$A$_{g}^-$ and $1^1$B$_{1g}^-$ all occur
{\it below} the allowed one-photon states as demonstrated also experimentally in Ref~\onlinecite{Aryanpour14a}. Finally, the density of the two-photon absorptions as well as their intensity profile
are very different from the noninteracting model. The weak intensities of the two-photon absorptions at the lowest energies, and much stronger
intensities at slightly higher energies are consequences of electron correlations \cite{McWilliams91a,FGuo94a}.
\begin{table*}
{\footnotesize
\caption{PPP-MRSDCI excited state energies for decapentaene and the PAH molecules of Figs.~\ref{molecules}(a)-(d) (in eV). The $^1$A$_g^-$ states in the last two columns have energies that are
nearly twice those of the lowest triplets T$_1$ and T$_2$.}
\label{t2}
\begin{tabular}{l}
\hline\hline 
\\ 
\hspace{2.7cm}one-photon states\hspace{0.85cm}2$^1$A$_{g}^-$\hspace{1.0cm}T$_1$\hspace{1.5cm}T$_2$\hspace{1.8cm}$^1$A$_{g}^-$ states ($\sim 2\times$T$_1$)\hspace{0.5cm}$^1$A$_{g}^-$ states ($\sim 2\times$T$_2$)  \\ \\ 
Decapentaene\hspace{0.92cm}4.12~(1$^1$B$_{u}^+$)\hspace{1.55cm}3.06\hspace{0.57cm}1.52~(1$^3$B$_u$)\hspace{0.55cm}2.45~(1$^3$A$_g$)\hspace{1.8cm}3.06~($2^{1}$A$_{g}^{-}$)\hspace{1.5cm}4.81~($3^{1}$A$_{g}^{-}$) \\ \\ 
Ovalene\hspace{1.67cm}2.91~(1$^1$B$_{2u}^+$)\hspace{1.5cm}3.03\hspace{0.55cm}1.57~(1$^3$B$_{2u}$)\hspace{0.45cm}2.65~(1$^3$B$_{3u}$)\hspace{1.65cm}3.03~($2^{1}$A$_{g}^{-}$)\hspace{1.5cm}5.27~($11^{1}$A$_{g}^{-}$)\\ \hspace{2.7cm}3.56~(1$^1$B$_{3u}^+$)\\ \hspace{2.7cm}4.18~(2$^1$B$_{2u}^+$) \\ \\
Coronene\hspace{1.47cm}4.11~(1$^1$B$_{2u}^+$, 1$^1$B$_{3u}^+$)\hspace{0.5cm}3.96\hspace{0.55cm}2.38~(1$^3$B$_{2u}$)\hspace{0.45cm}3.04~(1$^3$B$_{3u}$, 2$^3$B$_{2u}$)\hspace{0.6cm}4.77~($5^{1}$A$_{g}^{-}$)\hspace{1.55cm}6.07~($12^{1}$A$_{g}^{-}$) \\ \\
HBC\hspace{2.05cm}3.37~(1$^1$B$_{2u}^+$, 1$^1$B$_{3u}^+$)\hspace{0.5cm}3.30\hspace{0.55cm}2.07~(1$^3$B$_{2u}$)\hspace{0.45cm}2.72~(1$^3$B$_{3u}$, 2$^3$B$_{2u}$)\hspace{0.6cm}4.12~($5^{1}$A$_{g}^{-}$)\hspace{1.55cm}5.35~($16^{1}$A$_{g}^{-}$)\\ \\
Circumcoronene\hspace{0.6cm}2.66~(1$^1$B$_{2u}^+$, 1$^1$B$_{3u}^+$)\hspace{0.5cm}2.75\hspace{0.55cm}1.50~(1$^3$B$_{2u}$)\hspace{0.45cm}1.97~(1$^3$B$_{3u}$, 2$^3$B$_{2u}$)\hspace{0.6cm}2.94~($3^{1}$A$_{g}^{-}$)\hspace{1.55cm}3.68~($8^{1}$A$_{g}^{-}$) \\ \\
\hline \hline
\end{tabular}
}
\end{table*}
\par Fig.~\ref{Huckel}(d) presents similar analysis of the interacting spectra for ovalene.
The nondegeneracy of the H$\to$L lowest transition in the H\"uckel limit implies the absence of the 1$^1$B$_{2u}^-$ state here.
The degeneracy of the H$\to$L+1 and the H$-$1$\to$L transitions means however that the 1$^1$B$_{3u}^-$ dark state does exist,
as the superposition of the two excitations with relative minus sign between them.
As in the noninteracting model, three distinct allowed optical states are therefore found, $1^1$B$_{2u}^+$, $1^1$B$_{3u}^+$ and
$2^1$B$_{2u}^+$ with $y$-, $x$- and $y$- polarizations, respectively, where $x$ and $y$ are defined to be the longitudinal and transverse axes of
ovalene. The lowest two-photon state, the $2^1$A$_{g}^-$ is very slightly above the $1^1$B$_{2u}^+$ but significantly below the
dominant one-photon absorption 1$^1$B$_{3u}^+$. It is conceivable that the higher energy of the of the $2^{1}$A$_{g}^-$ here is a consequence of the size inconsistency inherent in the MRSDCI approach \cite{Cave90a}, given the very small energy difference we find between the 2$^1$A$_g^-$ and the 1$^1$B$_{2u}^+$. This, however, does not change our overall conclusions. For the lower symmetry and the consequent larger splittings between the $^1$A$_g^-$ and the
$^1$B$_{1g}^-$, we find that the overall density of the two-photon absorptions and the intensity profile is very similar to coronene.
\par Summarizing this subsection, electron correlations have a profound effect on the energetics of the two-photon states in both
the $D_{6h}$ PAHs and $D_{2h}$ ovalene in that the ordering of the excited states is substantially different from that expected within the
noninteracting H\"uckel model. This is expected to a large extent, and does not indicate any difference between the two classes of PAH molecules.
In the rest of the paper we point out that correlation effects in fact are significantly stronger in ovalene than in the $D_{6h}$ PAHs.
\par\noindent{\it Comparison, $D_{2h}$ versus polyenes and $D_{6h}$.} Table~\ref{t2} lists the energies of the lowest optical one-photon states, the
lowest two-photon 2$^1$A$_g^-$ state and the   
lowest two triplet states T$_1$ and T$_2$ for decapentaene and the PAH molecules of Figs.~\ref{molecules}(a)-(d). As mentioned in section I, several
of the low-lying $^1A_g$ states in polyenes are of two-triplet character. These earlier results \cite{Tavan87a} are reproduced in  
Table~\ref{t2}, where the energies of the 2$^1$A$_g^-$ and 3$^1$A$_g^-$ in decapentaene are twice those of T$_1$ and T$_2$, respectively.
These states have T$_1\otimes$T$_1$ and T$_2\otimes$T$_2$ characters, respectively \cite{Tavan87a} (we have not listed the T$_1\otimes$T$_2$ eigenstate 
which lies in the two-photon inactive B$_u^-$ subspace). 
We have determined that in the PAHs also there exist $^1$A$_g^-$ states, albeit
with quantum number greater than 2 in most cases, whose energies are nearly twice of T$_1$ and T$_2$, respectively. These $^1A_g$ states and their energies have also been listed in Table~\ref{t2}. 
\par The 2$^1$A$_g^-$ is not a two-triplet state in any of the $D_{6h}$ PAH molecules, for which even higher lying $^1$A$_g^-$
states have energies twice that of T$_1$ (small
deviations from exactly 2 $\times$ E(T$_1$) is expected as at the lowest energies in the $^1$A$_g^-$ subspace weak binding between the spin excitations
can occur \cite{Tavan87a}). Interestingly, the quantum number of the apparent T$_1\otimes$T$_1$ $^1$A$_g^-$ state decreases from 5 in coronene and HBC to
3 in circumcoronene, suggesting that in still larger $D_{6h}$ PAHs the 2$^1$A$_g$ may be the T$_1\otimes$T$_1$ state. This size dependence is consistent
with the size-dependent behavior of $^1$A$_g^-$ states noted previously in polyenes, where with increasing size the energy of the 2$^1$A$_g^-$ 
relative to that of the 1$^1$B$_u^+$
decreases, and also more and more $^1$A$_g^-$ states appear below the 1$^1$B$_u^+$. Table ~\ref{t2} also lists the $^1$A$_g^-$ states whose energies
are at nearly twice that of T$_2$. The quantum numbers of these states are even larger. That the 2$^1$A$_g^-$ is not T$_1\otimes$T$_1$, and that the
apparently T$_1\otimes$T$_1$ and T$_2\otimes$T$_2$ states are not consecutive $^1$A$_g^-$ states 
both indicate that correlation effects are weaker in the $D_{6h}$ PAHs  
than in polyenes.
\par Surprisingly, the 2$^1$A$_g^-$ is the T$_1\otimes$T$_1$ state in ovalene from energetics, indicating very strongly that as per our measures of
correlation effects discussed in section I, ovalene is more strongly correlated than the $D_{6h}$ PAH molecules. There is an apparent contradiction
here when one compares the one- and two-photon spectra in Fig.~2, where the 2$^1$A$_g^-$ occurs below the allowed one-photon state in coronene
but slightly above the one-photon state in ovalene. This is resolved when it is realized that the lowest one-photon allowed state in ovalene
actually corresponds to the {\it forbidden dark state} in coronene, which is located below the 2$^1$A$_g^-$. {\it The optical one-photon state
in coronene at 4.1 eV thus corresponds to the strongest peak in ovalene at $\sim$ 3.56 eV, and the 2$^1$A$_g^-$ in ovalene is therefore much lower in energy 
relative to the optical state in ovalene than in coronene.}
The apparent  T$_2\otimes$T$_2$
is again a $^1$A$_g^-$ state with high quantum number, which then indicates that correlation effects are weaker in ovalene than in polyenes.
Our conclusions regarding the ``degree'' of correlation effects in the three classes of systems of interest here is supported by the 
extensive wavefunction analyses we have performed, which we now discuss.
\par
\begin{figure*}
\includegraphics[width=6.0in]{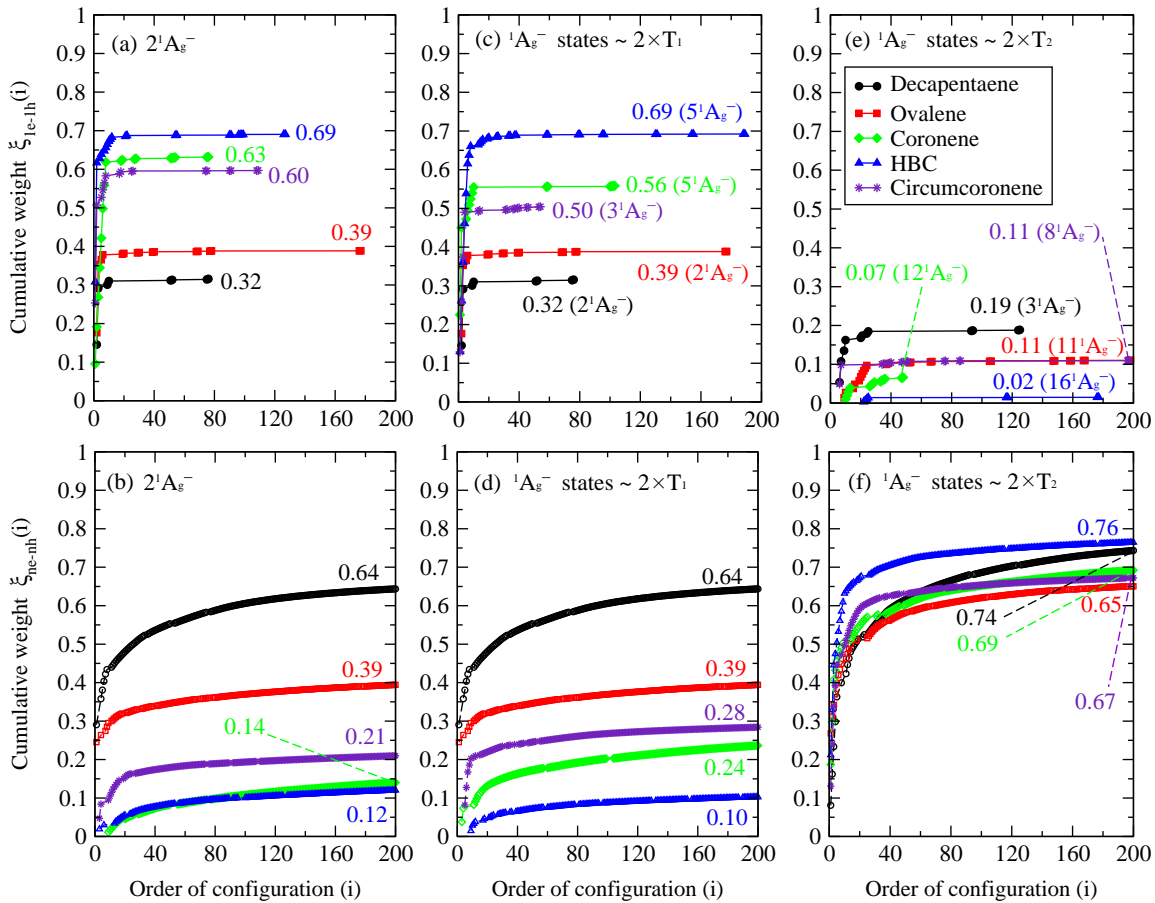}
\caption{Cumulative weight $\xi(i)$ versus configuration order $i$ for decapentaene and the PAH molecules of Figs.~\ref{molecules}(a)-(d). 
Top and bottom panels give the normalized cumulative weights of 1e-1h and ne-nh (n$\geq$2) excitations to $^1$A$_g^-$ eigenstates of interest (see Table~\ref{t2}).
Discrete points as appear on the curves in the top panel are not shown in the bottom panel, where the configurations do appear almost continuously.
The numbers against each curve correspond to the total relative weight for the eigenstate, at $i=200$.}

\label{wavefunctions}
\end{figure*}

 The MRSDCI many-electron wavefunctions are of the form $|\Psi\rangle = \sum_m \beta_m|m\rangle$, where $|m\rangle$ is an excitation from the HF ground state, and $\beta_m$ are their normalized coefficients.
In linear polyenes, the strongly correlated $^1$A$_g^-$ states have stronger contributions from multielectron ne-nh excitations (n $\geq$ 2) 
than from the 1e-1h excitations. This is actually a requirement for the state to be strictly two-triplet, since each triplet excitation is a 
1e-1h excitation within the MO picture. It is also taken to be a signature that the corresponding wavefunction in the VB language
is of covalent character. Because of the gigantic dimensions of the $^1$A$_g^-$ subspaces in 
Table~\ref{t1} approximation becomes necessary for similar analyses of eigenstates of the PAHs.
We define the {\it''cumulative relative weight''} $\xi_{a}(i)=\sum_{m=1}^{i}|\beta_{m}^{a}|^2$, where $a$ = 1e-1h or ne-nh, and the $\beta_{m}^{a}$ have been sorted in descending order with the largest term first. We then numerically compute
$\xi_{a}(i)$ as a function of $i$. The top and bottom panels in Fig.~\ref{wavefunctions} correspond to $\xi_{\mathrm{1e-1h}}(i)$ and $\xi_{\mathrm{ne-nh}}(i)$, respectively, for decapentaene and the PAH molecules of Figs.~\ref{molecules}(a)-(d). As seen in the figures,
rapid convergence is reached in $\xi_{a}(i)$ with increasing $i$, allowing characterizations of
wavefunctions as primarily 1e-1h or ne-nh. 
\begin{table*}
{\footnotesize
\caption{Dominant contributions to the  wavefunctions of T$_2$ and the $^{1}$A$_{g}^{-}$ states with energies $\sim$ 2$\times$E(T$_2$). 1e-1h
and 2e-2h excitations are denoted by one ($\rightarrow$) and two ($\rightrightarrows$) arrows, respectively.
Numbers in front of the excitations are the absolute values of their coefficients in the wavefunction.  \\}
\label{t3}
\begin{tabular}{l}
\hline\hline 
\\ 
Wavefunction:\hspace{1.5cm}T$_2$\hspace{2.8cm}$^1$A$_{g}^-$ states ($\sim$2$\times$T$_2$)  \\ \\ \hline\hline  \\ 
Ovalene\hspace{1.5cm}1$^3$B$_{3u}$:\hspace{3.25cm}$11^{1}$A$_{g}^{-}$:\\ \hspace{2.5cm}(0.56)~$|$H$\rightarrow$L$+1\rangle$\hspace{2.0cm}(0.52)~$|$H,H$-1\rightrightarrows$L,L$+1\rangle$\\\hspace{2.5cm}(0.56)~$|$H$-1\rightarrow$L$\rangle$\hspace{1.92cm}(0.21)~$|$H$\rightrightarrows$L$\rangle$\\\hspace{6.65cm}(0.17)~$|$H$-1\rightrightarrows$L$\rangle$\\\hspace{6.65cm}(0.17)~$|$H$\rightrightarrows$L$+1\rangle$ \\ \\
Coronene\hspace{1.28cm}1$^3$B$_{3u}$:\hspace{3.25cm}$12^{1}$A$_{g}^{-}$:\\ \hspace{2.5cm}(0.58)~$|$H$1\rightarrow$L$1\rangle$\hspace{2.0cm}(0.43)~$|$H$1$,H$2\rightrightarrows$L$1$,L$2\rangle$\\\hspace{2.5cm}(0.58)~$|$H$2\rightarrow$L$2\rangle$\hspace{2.0cm}(0.33)~$|$H$1\rightrightarrows$L$1\rangle$\\\hspace{2.52cm}$2^3$B$_{2u}$:\hspace{3.22cm}(0.33)~$|$H$2\rightrightarrows$L$2\rangle$\\\hspace{2.5cm}(0.58)~$|$H$1\rightarrow$L$2\rangle$\hspace{2.0cm}(0.16)~$|$H$1\rightrightarrows$L$2\rangle$\\\hspace{2.5cm}(0.58)~$|$H$2\rightarrow$L$1\rangle$\hspace{2.0cm}(0.16)~$|$H$2\rightrightarrows$L$1\rangle$ \\ \\
HBC\hspace{1.85cm}1$^3$B$_{3u}$:\hspace{3.25cm}$16^{1}$A$_{g}^{-}$:\\\hspace{2.5cm}(0.60)~$|$H$1\rightarrow$L$1\rangle$\hspace{2.0cm}(0.46)~$|$H$1$,H$2\rightrightarrows$L$1$,L$2\rangle$\\\hspace{2.5cm}(0.60)~$|$H$2\rightarrow$L$2\rangle$\hspace{2.0cm}(0.34)~$|$H$1\rightrightarrows$L$1\rangle$\\\hspace{2.52cm}2$^3$B$_{2u}$:\hspace{3.22cm}(0.34)~$|$H$2\rightrightarrows$L$2\rangle$\\\hspace{2.5cm}(0.60)~$|$H$1\rightarrow$L$2\rangle$\\\hspace{2.5cm}(0.60)~$|$H$2\rightarrow$L$1\rangle$ \\ \\
Circumcoronene\hspace{0.4cm}1$^3$B$_{3u}$:\hspace{3.25cm}$8^{1}$A$_{g}^{-}$:\\ \hspace{2.5cm}(0.59)~$|$H$1\rightarrow$L$1\rangle$\hspace{2.0cm}(0.36)~$|$H$1$,H$2\rightrightarrows$L$1$,L$2\rangle$\\\hspace{2.5cm}(0.59)~$|$H$2\rightarrow$L$2\rangle$\hspace{2.0cm}(0.31)~$|$H$1\rightrightarrows$L$1\rangle$\\\hspace{2.52cm}2$^3$B$_{2u}$:\hspace{3.22cm}(0.33)~$|$H$2\rightrightarrows$L$2\rangle$\\\hspace{2.5cm}(0.59)~$|$H$1\rightarrow$L$2\rangle$\\\hspace{2.5cm}(0.59)~$|$H$2\rightarrow$L$1\rangle$ \\ \\
\hline \hline

\end{tabular}
}
\end{table*}
\par In Figs.~\ref{wavefunctions}(a)-(b) for the 2$^1$A$_g^-$ state in decapentaene, $\xi_{\mathrm{ne-nh}}(i)$ is already 0.64 at $i=200$, while 
$\xi_{\mathrm{1e-1h}}(i)$ has nearly converged to 0.32 at $i=76$. Figs.~\ref{wavefunctions}(e)-(f) for 3$^1$A$_g^-$ state in decapentaene show even stronger 
contribution of ne-nh (0.74) against 1e-1h (0.19) excitations. Note that 1e-1h configurations become more improbable with increasing $i$ and 
therefore $\xi_{\mathrm{1e-1h}}$ can be taken as the converged value at $i\sim100$. The results of Fig.~\ref{wavefunctions} 
therefore indicate the strongly correlated characters of the  2$^1$A$_g^-$ and 3$^1$A$_g^-$ in decapentaene, in agreement with earlier characterizations of these
eigenstates in polyenes.
\par In Figs.~\ref{wavefunctions}(a)-(b), $\xi_{\mathrm{1e-1h}}(i)$ for the 2$^1$A$_g^-$ states of the $D_{6h}$ PAH molecules have nearly converged to values 
equal to or larger than 0.6. The corresponding $\xi_{\mathrm{ne-nh}}$ remain at or below 0.21 even at $i=200$, indicating once again
weaker correlation effects than in the polyenes \cite{Aryanpour14a}. Somewhat surprisingly, $^1$A$_g^-$ states identified as T$_1\otimes$T$_1$ in the
$D_{6h}$ PAH molecules from energetics also have larger contributions from 1e-1h excitations than from ne-nh excitations (see Fig.~\ref{wavefunctions}(c)-(d)).
The strong dominance of 1e-1h over ne-nh excitations in the T$_1\otimes$T$_1$ state is a novel aspect of $D_{6h}$ PAHs. This is in apparent agreement with 
our previous observation that based on the proximity between the T$_1$ state and the ionic one-photon optical states,
the T$_1$ state in the $D_{6h}$ PAHs are of ionic character within the VB language \cite{Aryanpour14a}.
Thus neither the T$_1$ nor the T$_1\otimes$T$_1$ states in the $D_{6h}$ PAHs can be considered as covalent in these molecules. 
We have mentioned above that in the $D_{6h}$ PAHs larger than circumcoronene it is likely that the 2$^1$A$_g^-$ state is the T$_1\otimes$T$_1$. Since the
number of covalent VB diagrams increases rapidly with size \cite{Soos84a,Ramasesha84a}, it is conceivable that in these cases the 2$^1$A$_g^-$ might acquire covalent character.
\par In contrast to the $D_{6h}$ PAHs, the 2$^1$A$_g^-$ wavefunction in ovalene 
has nearly equal contributions from
1e-1h and ne-nh excitations, as seen in Figs.~\ref{wavefunctions}(a)-(b), where $\xi_{\mathrm{1e-1h}}(i=177)=\xi_{\mathrm{ne-nh}}(i=200)=0.39$. 
Thus both from energetic considerations (Table~\ref{t2}) and from wavefunction analyses, we reach the same conclusion, viz., 
correlation effects are far stronger in $D_{2h}$
ovalene than in any of the $D_{6h}$ molecules in Fig.~\ref{molecules}.
\par The $^1$A$_g^-$ states with energies $\sim 2\times$E(T$_2$) in all the PAHs are at much higher energies with respect to the lowest one-photon allowed
states, in contrast to decapentaene. As seen in 
Figs.~\ref{wavefunctions}(e)-(f), these $^1$A$_g^-$ states are very strongly ne-nh in character as $\xi_{\mathrm{ne-nh}}(i=200)\sim0.67-0.76$ with $\xi_{\mathrm{1e-1h}}<0.2$. In Table~\ref{t3} we have given the leading many-electron configurations that dominate the wavefunctions of T$_2$ and the
$^1$A$_g^-$ states with energies $\sim 2\times$E(T$_2$) for all PAH molecules of Fig.~\ref{molecules}. Comparison of these wavefunction components
shows clearly that these high energy  $^1$A$_g^-$ states can indeed be identified as T$_2\otimes$T$_2$. 
In between the nominally T$_1\otimes$T$_1$ and T$_2\otimes$T$_2$ $^1$A$_g$ states in the PAHs
there occur several additional $^1$A$_g^-$ states which are difficult to characterize in simple terms. Earlier work by
Chandross and Mazumdar \cite{Chandross99a} on linear polyenes had shown that two-photon states here can be classified as excitations that are
predominantly 2e-2h triplet-triplet at the lowest energies, predominantly 1e-1h charge-transfer at intermediate energies and predominantly 
2e-2h singlet-singlet at the highest energies. The occurrence of the T$_2\otimes$T$_2$ states at very high energies in the PAHs, along with the 
mixed character of the $^1$A$_g$ states below this state show that while electron correlations still split the 2e-2h excited states into 
triplet-triplet and singlet-singlet, their energetic locations are quite different from those in polyenes.
\section{Conclusions}
\label{sec:conclude}
\par In summary, we have performed accurate high order CI calculations of spin-singlet one- and two-photon excited states, and spin-triplet states for a number of 2D
PAH molecules that indicate the strong role of electron correlations in these systems. The predicted excited state ordering in all cases is significantly
different from the predictions of one-electron theory. In the past, theoretical work on 1D molecules, oligomers and polymers have been accompanied by
extensive experimental work \cite{Sariciftci97a,Vardeny09a,Jorio07a,Hudson82a,Kohler88a,Lawrence94a}. It is hoped that these first results for the PAH molecules
will similarly motivate spectroscopist to study their singlet and triplet energy spectra. As pointed out in our earlier work on the $D_{6h}$ PAH
molecules \cite{Aryanpour14a}, the relative ordering of one- versus two-photon states at the 
lowest energies suggests that correlation effects might be important even in the thermodynamic limit of graphene and gives us motivation to pursue theoretical and 
experimental investigation of even larger graphene fragments \cite{Yan11a}.  
\par We defined two different semiquantitative measures of correlation effects here, and from both counts
the $D_{2h}$ molecule ovalene is significantly more strongly correlated than the three $D_{6h}$ PAH molecules we investigated. It is tempting to ascribe the difference to ovalene being the ``most quasi-1D'', but the overall
one-electron bandwidths of the molecules are quite comparable, ranging from  5.35$|t|$ in coronene to 5.68$|t|$ in circumcoronene, increasing slowly
with system size. Indeed, larger correlation effect in ovalene with 32 C-atoms than in circumcoronene with 54 C-atoms is quite unexpected, given the
very rapid increase in the number of covalent VB diagrams with size. One obvious difference between the two classes is the degeneracies of the one-electron frontier MOs in the $D_{6h}$ molecules
versus their nondegeneracy in $D_{2h}$. The characters of the frontier MOs in ovalene are however very different from that in the polyenes, in that
the transitions H $\to$ L$+$1 and H$-$1 $\to$ L, forbidden in the polyenes, are allowed in ovalene. Thus the nondegeneracy in ovalene results from weak
symmetry breaking relative to the  $D_{6h}$ PAHs, and does not result in perfect alternation of one-electron levels with opposite symmetries,
as occurs in polyenes.
Our calculated results here suggest much larger covalent character of the ground state itself in the less symmetric case. There exist a variety of
many-body techniques for studying the ground state wavefunctions of large correlated-electron systems. We are currently pursuing such a study, using the 
path integral renormalization group technique used previously by one of us \cite{Dayal12a}.
\section*{ACKNOWLEDGMENTS}
This work was supported by NSF grant No. CHE-1151475 and by the Indo-US Science and Technology Forum Award 37-2012/2013-14.
\end{document}